\documentclass[journal]{IEEEtran}
\usepackage{cite}
\usepackage{amsmath}
\usepackage{amsfonts}
\usepackage{bm}
\usepackage{extarrows}
\usepackage{graphicx}
\usepackage[caption=false]{subfig}
\usepackage{xcolor}
\usepackage{esvect}

\graphicspath{{figure/}}

\newcommand{\mr}{\mathrm}

\newtheorem{proposition}{Proposition}

\newtheorem{corollary}{Corollary}
\begin{document}

\title{On the Performance of Turbo Signal Recovery\\
with Partial DFT Sensing Matrices
}

\author{Junjie~Ma,
        ~Xiaojun~Yuan,~\IEEEmembership{Member,~IEEE}
        and~Li~Ping,~\IEEEmembership{Fellow,~IEEE}

 \thanks{J.~Ma and Li Ping are with the Department of Electronic Engineering, City University of Hong Kong, Hong Kong, SAR, China. (e-mail: junjiema2-c@my.cityu.edu.hk; eeliping@cityu.edu.hk.) The work of J.~Ma and Li Ping is supported by a grant from the Research Grant Council of the Hong Kong SAR, China [Project No. AoE/E-02/08 and CityU 118013]. X. Yuan is with the School of Information Science and Technology, ShanghaiTech University (email: yuanxj@shanghaitech.edu.cn). The work of X.~Yuan was supported in part by the National Natural Science Foundation of China under Grant 61471241.}
}
\maketitle

\begin{abstract}
This letter is on the performance of the turbo signal recovery (TSR) algorithm for partial discrete Fourier transform (DFT) matrices based compressed sensing. Based on state evolution analysis, we prove that TSR with a partial DFT sensing matrix outperforms the well-known approximate message passing (AMP) algorithm with an independent identically distributed (IID) sensing matrix.
\end{abstract}

\begin{IEEEkeywords}
Turbo compressed sensing, signal recovery, AMP, partial DFT, state evolution.
\end{IEEEkeywords}

\IEEEpeerreviewmaketitle

\section{Introduction}
The approximate message passing (AMP) algorithm \cite{Donoho2009,Bayati2011,Rangan2011,Kamilov2014,krzakala2012,Donoho2012,Phil2013,Tan2014,Guo2014} is an efficient signal recovery method for compressed sensing. Its convergence is asymptotically guaranteed for sensing matrices with independent identically distributed (IID) entries using the state evolution technique \cite{Donoho2009,Bayati2011}. The fixed points of the state evolution for AMP\footnote{Throughout this letter, AMP refers to AMP-MMSE \cite{Reeves2012}.} include the optimal minimum mean squared-error estimation (MMSE) solution \cite{Guo2005Replica,Guo2009,Tulino2013}. This indicates that AMP is asymptotically optimal when the state evolution equation has a unique solution.\par
AMP can also be applied to problems involving non-IID sensing matrices \cite{Barbier2013,Rangan2014}. However, the state evolution technique is not directly applicable in this case.\par
Alternative techniques have been developed for non-IID sensing matrices \cite{Kabashima2014,Cakmak2014,Ma2015}. It has been observed that these techniques with partial discrete Fourier transform (DFT) matrices \cite{Kit2014,Vehkapera2014,Wen2014,Oymak2014} can outperform AMP with IID sensing matrices under proper normalization conditions. The comparisons in \cite{Kabashima2014,Cakmak2014,Ma2015} were based on simulations and no analytical results have been reported so far.\par
{{This letter is on the performance analysis of turbo signal recovery (TSR) with a partial DFT sensing matrix \cite{Ma2015}. We prove based on state evolution that TSR with a partial DFT matrix (TSR-DFT) outperforms AMP with an IID Gaussian matrix (AMP-IID). Since the state evolution technique for AMP does not apply to problems involving a partial DFT matrix (AMP-DFT), we compare TSR-DFT and AMP-DFT through simulations. Our numerical results suggest that TSR-DFT converges faster than AMP-DFT.}}
\section{Problem Description}
Consider the following linear system:
\begin{equation}\label{Equ:model}
\bm{y}=\bm{F}_{\mathrm{partial}}\bm{x}+\bm{n}
\end{equation}
where $\bm{x}\in\mathbb{C}^{N\times1}$ is a sparse signal, $\bm{n}\sim\mathcal{CN}(\mathbf{0},\sigma^2\bm{I})$ the additive white Gaussian noise (AWGN) and $\bm{F}_{\mathrm{partial}}\in\mathbb{C}^{M\times N}$ ($M<N$) a partial DFT matrix consisting of $M$ randomly selected rows of the normalized DFT matrix $\bm{F}$. The $(m,n)$th entry of $\bm{F}$ is given by $\frac{1}{\sqrt{N}}\exp\big(-\mr{j}\frac{2\pi(m-1)(n-1)}{N}\big)$. \par
We assume that the entries of $\bm{x}$ are IID. The $j$th entry $x_j$ follows the Bernoulli-Gaussian distribution \cite{Rangan2011}
\begin{equation}\label{Equ:BG}
x_j \sim
\begin{cases}
0 & \text{probability} = 1 - \lambda,\\
\mathcal{CN}(0,\lambda^{-1}) & \text{probability} = \lambda.
\end{cases}
\end{equation}
By this definition, $\mr{E}[|x_j|^2] = 1$.  Here $\lambda$ determines the sparsity of the system. The partial DFT matrix can be rewritten as
\begin{equation}
\bm{F}_{\mathrm{partial}}=\bm{SF},
\end{equation}
where $\bm{S}$ consists of $M$ randomly selected rows of the identity matrix. We define the following auxiliary vector:
\begin{equation}\label{Equ:u_def}
\bm{z}=\bm{Fx}.
\end{equation}
Combining \eqref{Equ:model} and \eqref{Equ:u_def}, we have
\begin{equation}\label{Equ:model_u}
\bm{y}=\bm{S}\bm{z}+\bm{n}.
\end{equation}
Our objective is to recover $\bm{x}$ based on $\bm{y}$ under the assumption that $\bm{x}$ is sparse with $\lambda <1$.
 \begin{figure}[htbp]
\centering
\includegraphics[width=.48\textwidth]{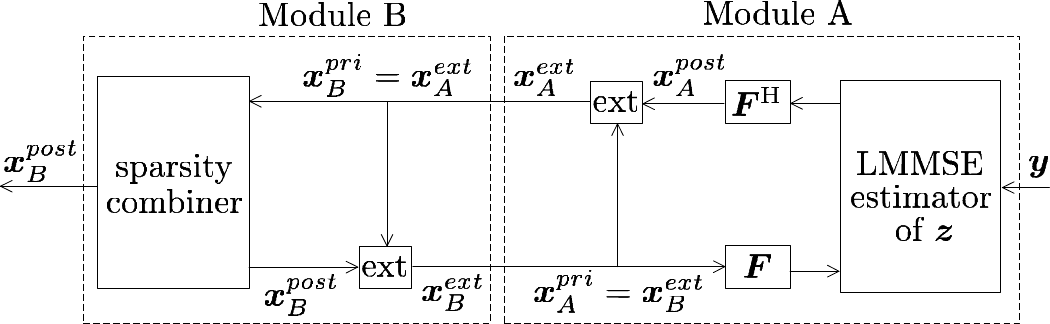}
\caption{Functional diagram of a standard turbo processor.}\label{Fig:standard}
\end{figure}
\section{Turbo Signal Recovery}
 \begin{figure*}[!t]
\centering
\includegraphics[width=.75\textwidth]{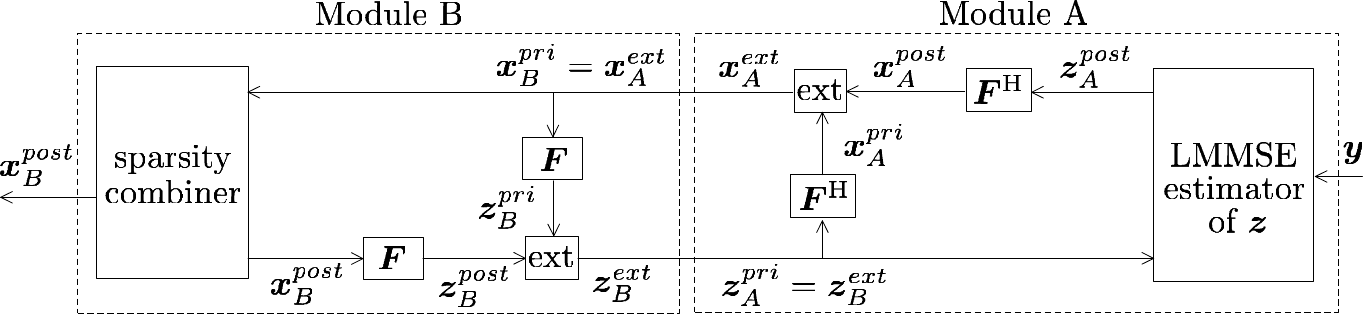}
\caption{Functional diagram of the turbo signal recovery (TSR) algorithm \cite{Ma2015}.``ext" represents extrinsic message computation.}\label{Fig:model}
\end{figure*}
\subsection{Standard Turbo Processor}
{{Fig.~\ref{Fig:standard} shows a standard turbo-type signal processor \cite{Berrou1993} for the problem under consideration. The related operations can be grouped into two modules labeled as A and B. Module A  is a linear minimum mean-squared error (LMMSE) estimator \cite{Kay1993} of $\bm{x}$ based on \eqref{Equ:model} \textit{without} the sparsity information, while module B estimates $\bm{z}$ \textit{based on} the sparsity information in \eqref{Equ:BG}. The two modules work iteratively. \par
 Since LMMSE estimation is standard, we will focus on module B. The input of module B, denoted by $\bm{x}_B^{pri}$ (see Fig.~\ref{Fig:standard}), is modeled as \cite{Ma2015}
  \begin{equation}\label{Equ:AWGN-revise}
  \bm{x}_B^{pri} = \bm{x}+\bm{w},
  \end{equation}
  where $\bm{w}$ is IID Gaussian and independent of $\bm{x}$. For each $j$, the sparsity combiner produces the \textit{a posteriori} mean $\mr{E}_{x_j}\{x_j|\bm{x}_{B}^{pri}\}$ based on the AWGN assumption in \eqref{Equ:AWGN-revise} and the sparsity constraint in \eqref{Equ:BG}. Let ``$_{\sim j}$'' denote indices excluding $j$. The extrinsic mean is defined as $\mr{E}_{x_j}\{x_j|\bm{x}_{B,\sim j}^{pri}\}$. Since $\bm{x}_{B}^{pri}$ is assumed to be an AWGN observation of $\bm{x}$, the extrinsic mean will not improve during the iterative process based on Fig.~\ref{Fig:standard}. The problem here is that the sparsity constraint is symbol-by-symbol and so $\bm{x}_{B,\sim j}^{pri}$ does not provide any information about $x_j$. For details, see \cite{Ma2015}.}}
\subsection{Turbo Signal Recovery}
 The TSR algorithm proposed in \cite{Ma2015} is listed in Algorithm 1 and graphically illustrated in Fig.~\ref{Fig:model}. The TSR algorithm computes the extrinsic message of $\bm{z}$ (instead of $\bm{x}$) for module B. This avoids the above mentioned problem for the standard turbo processor. Refer to \cite{Ma2015} for more details.
\noindent\\[5pt]
    \rule{.48\textwidth}{0.7pt}\\
    \textbf{Algorithm 1:} Turbo Signal Recovery (TSR)
    \\[-5pt]
    \rule{.48\textwidth}{0.4pt}
    \\
    \textbf{Initialization}: $\bm{z}^{pri}_A\leftarrow\mathbf{0}$, and $v^{pri}_A\leftarrow1$.\\
    \textbf{for} $iteration=1:T_{max}$\\
    1) Update
    \begin{equation}\label{Equ:x_A_pri}
    \bm{x}_A^{pri}\leftarrow\bm{F}^{\mr{H}}\bm{z}_A^{pri}.
    \end{equation}
    2) Compute the \textit{a posteriori} mean/variance of $\bm{z}$
    \begin{subequations}
    \begin{align}
    {\bm{z}}_A^{post} & \leftarrow {\bm{z}}_A^{pri}  + \frac{{v_A^{pri} }}{{v_A^{pri}  + \sigma ^2 }}{\bm{S}}^{\rm{H}} \left( {{\bm{y}} - {\bm{Sz}}_A^{pri} } \right), \\
    v_{A,j}^{post}  & \leftarrow v_A^{pri}  - \frac{{\left( {v_A^{pri} } \right)^2 }}{{v_A^{pri}  + \sigma ^2 }}\left( {{\bm{S}}^{\rm{H}} {\bm{S}}} \right)_{\left( {j,j} \right)} ,
    \end{align}
    \end{subequations}
    where $\left(\bm{S}^{\mr{H}}\bm{S}\right)_{(j,j)}$ denotes the $(j,j)$th entry of $\bm{S}^{\mr{H}}\bm{S}$.\\
    3) Compute the \textit{a posteriori} mean/variance of $\bm{x}$
    \begin{subequations}
    \begin{align}
    {\bm{x}}_A^{post} & \leftarrow {\bm{F}}^{\rm{H}} {\bm{z}}_A^{post} ,\\
    v_A^{post}  & \leftarrow \frac{1}{N}\sum\limits_{j = 1}^N {v_{A,j}^{post} }  = v_A^{pri}  - \frac{M}{N}\frac{{\left( {v_A^{pri} } \right)^2 }}{{v_A^{pri}  + \sigma ^2 }}.
    \end{align}
    \end{subequations}
    4) Compute the extrinsic mean/variance of $\bm{x}$
    \begin{subequations}
    \begin{align}
    v_B^{pri}  & \leftarrow v_A^{ext}  \leftarrow \left( {\frac{1}{{v_A^{post} }} - \frac{1}{{v_A^{pri} }}} \right)^{ - 1} , \label{Equ:v_A_ext}\\
    {\bm{x}}_B^{pri} & \leftarrow {\bm{x}}_A^{ext}  \leftarrow v_A^{ext} \left( {\frac{{{\bm{x}}_A^{post} }}{{v_A^{post} }} - \frac{{{\bm{x}}_A^{pri} }}{{v_A^{pri} }}} \right). \label{Equ:x_A_ext}
    \end{align}
    \end{subequations}
    5) Update
    \begin{equation}
    {\bm{z}}_B^{pri}  \leftarrow {\bm{Fx}}_B^{pri} .
    \end{equation}
    6) Compute the \textit{a posteriori} mean/variance of each $x_j$
    \begin{subequations}
    \begin{align}
    x_{B,j}^{post}  & \leftarrow {\rm{E}}_{x_j}\left\{ {x_j |x_{B,j}^{pri} } \right\}{\rm{,}} \\
    v_{B,j}^{post}  & \leftarrow {\rm{var}}_{x_j}\left\{ {x_j |x_{B,j}^{pri} } \right\}.
    \end{align}
    \end{subequations}
    7) Compute the \textit{a posteriori} mean/variance of $\bm{z}$
    \begin{subequations}
    \begin{align}
    {\bm{z}}_B^{post}  & \leftarrow {\bm{Fx}}_B^{post} , \\
    v_B^{post} & \leftarrow \frac{1}{N}\sum\limits_{j = 1}^N {v_{B,j}^{post} } .
    \end{align}
    \end{subequations}
    8) Compute the extrinsic mean/variance of $\bm{z}$
    \begin{subequations}\label{Equ:B_ext}
    \begin{align}
    v_A^{pri}  & \leftarrow v_B^{ext}  \leftarrow \left( {\frac{1}{{v_B^{post} }} - \frac{1}{{v_B^{pri} }}} \right)^{ - 1} , \label{Equ:v_B_ext}\\
    {\bm{z}}_A^{pri} & \leftarrow {\bm{z}}_B^{ext}  \leftarrow v_B^{ext} \left( {\frac{{{\bm{z}}_B^{post} }}{{v_B^{post} }} - \frac{{{\bm{z}}_B^{pri} }}{{v_B^{pri} }}} \right). \label{Equ:z_B_ext}
    \end{align}
    \end{subequations}
    \textbf{end}
    \\[-5pt]
    \rule{.48\textwidth}{0.7pt}
\section{State Evolution Analysis}
In the following, we analyze the state evolution of TSR-DFT \cite{Ma2015}, based on which we prove that TSR-DFT outperforms AMP-IID.
\subsection{MMSE Properties for an AWGN System}
Assume that $x$ has zero mean and unit variance. Consider the following observation of $x$ corrupted by AWGN,
\begin{equation}\label{Equ:AWGN}
r=x+w,
\end{equation}
where $w\sim\mathcal{CN}(0,\eta^{-1})$ is independent of $x$ and $\eta$ is the signal-to-noise ratio (SNR). Following \cite{GuoDongNing2011}, define
\begin{equation}\label{Equ:var}
\mathrm{var}_x\{x|r\}\equiv\mr{E}_x\Big\{ \big|x -\mr{E}_x\{x|r\}\big|^2 \big| r\Big\},
\end{equation}
and
\begin{equation}\label{Equ:mmse}
\mathrm{mmse}(\eta)\equiv\mr{E}_r\big\{ \mr{var}_x\{x|r\}\big\}.
\end{equation}
The following properties of $\mr{mmse}(\cdot)$ are due to \cite[Propositions 4 and 9]{GuoDongNing2011}:
\begin{subequations}
\begin{align}
\underline{\mr{Property 1:}}\quad&\mr{mmse}(\eta)\le\frac{1}{\eta}, \label{Equ:P1}\\
\underline{\mr{Property 2:}}\quad&\frac{\mr{d}\,\mr{mmse}(\eta)}{\mr{d}\,\eta}=-\mr{E}_r\Big\{ \big(\mr{var}_x\{x|r\}\big)^2\Big\}. \label{Equ:P2}
\end{align}
\end{subequations}
The above two properties are useful to our later discussions.
\subsection{State Evolution of TSR-DFT}
We use the \textit{a priori} variances $v_A^{pri}$ and $v_B^{pri}$ to measure the reliabilities of $\bm{x}_A^{pri}$ in \eqref{Equ:x_A_pri} and $\bm{x}_B^{pri}$ in \eqref{Equ:x_A_ext}, respectively. Our basic assumption is that $\bm{x}_B^{pri}$ in \eqref{Equ:x_A_ext} is an AWGN observation of $\bm{x}$:
\begin{equation}\label{Equ:AWGN_2}
\bm{x}_B^{pri} = \bm{x}+\bm{w},
\end{equation}
where $\bm{w}\sim\mathcal{CN}(\mathbf{0},v_B^{pri}\bm{I})$ is independent of $\bm{x}$.\par
Define
\begin{equation}\label{Equ:def_1}
v\equiv v_A^{pri} \text{ and } \eta\equiv\frac{1}{v_B^{pri}}.
\end{equation}
It is shown in \cite{Ma2015} that the state evolution equations of TSR are given by
\begin{subequations} \label{Equ:SE}
\begin{align}
\eta^{t+1}\equiv\phi(v^t)&=\frac{1}{\frac{N-M}{M}\cdot v^t +\frac{N}{M}\cdot\sigma^2}, \label{Equ:phi} \\
v^{t+1}\equiv\psi(\eta^{t+1})&=\left( \frac{1}{\mr{mmse}(\eta^{t+1})} - \eta^{t+1}\right)^{-1}, \label{Equ:psi}
\end{align}
\end{subequations}
where the superscripts represent the iteration indices, with initialization $v^0=1$.
\subsection{Convergence of State Evolution for TSR-DFT}
\begin{proposition}
$\phi(\cdot)$ and $\psi(\cdot)$ in \eqref{Equ:SE} are non-increasing functions.
\end{proposition}
\begin{IEEEproof}
It is straightforward to see that $\phi(\cdot)$ in \eqref{Equ:phi} is a non-increasing function of $v^t$. We now rewrite \eqref{Equ:psi} as $\psi(\eta^{t+1})=[f(\eta^{t+1})]^{-1}$, where
\begin{equation}
f\left( {\eta ^{t + 1} } \right) \equiv \frac{1}{{\mr{mmse}\left( {\eta ^{t + 1} } \right)}} - \eta ^{t + 1} .
\end{equation}
So,
\begin{equation}
\frac{{{\rm{d}\,}f\left( {\eta ^{t + 1} } \right)}}{{{\rm{d}}\,\eta ^{t + 1} }} = {\rm{ - }}\frac{1}{{\left[ {\mr{mmse}\left( {\eta ^{t + 1} } \right)} \right]^{\rm{2}} }}\frac{{{\rm{d\,}}\mr{mmse}\left( {\eta ^{t + 1} } \right)}}{{{\rm{d\,}}\eta ^{t + 1} }} - {\rm{1}}{\rm{.}}
\end{equation}
From Property 2 in \eqref{Equ:P2}, we have
\begin{equation}\label{Equ:derivative}
\frac{{{\rm{d}}f\left( {\eta ^{t + 1} } \right)}}{{{\rm{d}}\eta ^{t + 1} }} = \frac{{{\rm{E}}_r\big\{ {\left( {{\mathop{\rm var}} } \right)^2 } \big\} - \left( {{\rm{E}}_r\left\{ {{\mathop{\rm var}} } \right\}} \right)^2 }}{{\left( {{\rm{E}}_r\left\{ {{\mathop{\rm var}} } \right\}} \right)^2 }} \ge 0
\end{equation}
where $\mr{var}\equiv\mr{var}_x\left\{x|r=x+w\right\}$ and $w\sim\mathcal{CN}(0,\eta^{t+1})$. From \eqref{Equ:derivative}, $f(\cdot)$ is a non-decreasing function and so $\psi(\cdot)=[f(\cdot)]^{-1}$ is a non-increasing function.
\end{IEEEproof}\par\vspace{5pt}
Based on the monotonicity of the state transfer functions $\phi(\cdot)$ and $\psi(\cdot)$, it can be proved that $\{v^t\}$ and $\{\eta^{t+1}\}$ are monotone, i.e.,
\begin{equation}\label{Equ:monotone}
v^0  \ge v^1  \ge  \cdots  \ge v^\infty \text{ and } \eta ^1  \le \eta ^2  \le  \cdots  \le \eta ^\infty.
\end{equation}
In the first iteration, $t = 0$ in \eqref{Equ:phi} so
\begin{equation}\label{Equ:init}
v^0  = 1{\text{ and }}\eta ^1  = \left( {\frac{{N - M}}{M} + \frac{N}{M} \cdot \sigma ^2 } \right)^{ - 1}  > 0.
\end{equation}
Applying Property 1 in \eqref{Equ:P1} to \eqref{Equ:psi} yields
\begin{equation}\label{Equ:v_lower}
v^\infty   = \left( {\frac{1}{{\mr{mmse}\left( {\eta ^\infty  } \right)}} - \eta ^\infty  } \right)^{ - 1}  \ge 0.
\end{equation}
Combining \eqref{Equ:v_lower} and \eqref{Equ:phi}, we have
\begin{equation}\label{Equ:eta_upper}
\eta ^\infty   \le \left( {\frac{N}{M} \cdot \sigma ^2 } \right)^{ - 1} .
\end{equation}
Finally, from \eqref{Equ:monotone} and \eqref{Equ:v_lower}-\eqref{Equ:eta_upper}, we get
\begin{subequations}\label{Equ:bounds}
\begin{align}
1 &= v^0  \ge  \cdots  \ge v^\infty   \ge {\rm{0 ,}}\\
0 &< \eta ^1  \le  \cdots  \le \eta ^\infty   \le \frac{M}{N} \cdot \frac{1}{{\sigma ^2 }}.
\end{align}
\end{subequations}
\par
From \eqref{Equ:bounds}, the state sequences $\{v^t\}$ and $\{\eta^{t+1}\}$ are monotonic and bounded, and so they converge. Combining \eqref{Equ:phi} and \eqref{Equ:psi}, the stationary value $\eta^{\infty}$ is the solution of the following equation \cite{Ma2015}:
\begin{equation}\label{Equ:fixed}
\eta ^\infty   = \frac{{\mr{mmse} + \sigma ^2  - \sqrt {\left( {\mr{mmse} + \sigma ^2 } \right)^2  - 4\sigma ^2 \cdot\mr{mmse}\cdot\frac{M}{N}} }}{{2 \cdot \sigma ^2  \cdot \mr{mmse}}},
\end{equation}
where $\mr{mmse}$ is an abbreviation for $\mr{mmse}(\eta^{\infty})$. Note that \eqref{Equ:fixed} is consistent with the optimal MMSE performance obtained by the replica method. See \cite[Eqns.~(17) and (37)]{Tulino2013}.
\subsection{Comparison of TSR-DFT and AMP-IID}
Refer to the discussions in the Introduction. We now compare TSR-DFT and AMP-IID based on their state evolution equations. \par
The state evolution of AMP-IID is given by \cite[Eqn.~(41)]{Reeves2012},\cite[Eqns.~(18) and (20)]{Kit2014}\footnote{Note that the variances of the entries of the IID Gaussian matrix are $1/N$, instead of $1/M$ as assumed in \cite{Donoho2009,Bayati2011,Rangan2011}. This is for the convenience of comparison with TSR-DFT.}.
\begin{subequations}\label{Equ:SE_AMP}
\begin{align}
\eta _{{\text{AMP-IID}}}^{t + 1}  &= \frac{1}{{\frac{N}{M} \cdot v_{{\text{AMP-IID}}}^t  + \frac{N}{M} \cdot \sigma ^2 }}, \label{Equ:SE-AMP1}\\
v_{{\text{AMP-IID}}}^{t + 1}  &= \mathrm{mmse}\left( {\eta _{{\text{AMP-IID}}}^{t + 1} } \right), \label{Equ:SE-AMP2}
\end{align}
\end{subequations}
with initiation $v^{0}_{\text{AMP-IID}}=1$. \par
For TSR-DFT, we rewrite \eqref{Equ:SE} as
\begin{subequations}\label{Equ:SE_2}
\begin{align}
\eta _{{\text{TSR-DFT}}}^{t + 1}  &= \frac{1}{{\frac{N}{M} \cdot v_{{\text{TSR-DFT}}}^t  + \frac{N}{M} \cdot \sigma ^2 }}, \label{Equ:SE2-a}\\
\frac{N}{{N - M}} \cdot v_{{\text{TSR-DFT}}}^{t + 1}  &= \left( {\frac{1}{{\mr{mmse}\left( {\eta _{{\text{TSR-DFT}}}^{t + 1}} \right)}} - \eta _{{\text{TSR-DFT}}}^{t + 1} } \right)^{ - 1}. \label{Equ:SE2-b}
\end{align}
\end{subequations}
The following helps to see the equivalence of \eqref{Equ:SE} and \eqref{Equ:SE_2}:
\begin{equation}\label{Equ:norm}
\frac{N}{{N - M}} \cdot v_{{\text{TSR-DFT}}}^t  \equiv v^t {\text{ and  }}\eta _{{\text{TSR-DFT}}}^{t + 1}  \equiv \eta ^{t + 1} ,\ \forall t.
\end{equation}
A factor of $N/(N-M)$ is used \eqref{Equ:norm} to match \eqref{Equ:SE2-a} with \eqref{Equ:SE-AMP1}, which facilitates the proof of the proposition below.\vspace{5pt}
\begin{proposition}
$v_{{\text{TSR-DFT}}}^t  \le v_{{\text{AMP-IID}}}^t,\text{ for } t\ge0.$
\end{proposition}\vspace{5pt}
\begin{IEEEproof}
We prove by induction on $t$. The initial conditions are $v_{\text{AMP-IID}}^{0}=1$ and $v^{0}=1$. So from \eqref{Equ:norm},
\begin{equation}
v_{{\text{TSR-DFT}}}^0  = \frac{{N - M}}{N} \cdot v^0  = \frac{{N - M}}{N} < v_{{\text{AMP-IID}}}^0 .
\end{equation}
Now suppose
\begin{equation}\label{Equ:induct_assume}
v_{{\text{TSR-DFT}}}^t  \le v_{{\text{AMP-IID}}}^t .
\end{equation}
It suffices to prove that
\begin{equation}\label{Equ:induct_proof}
v_{{\text{TSR-DFT}}}^{t + 1}  \le v_{{\text{AMP-IID}}}^{t + 1} .
\end{equation}
Combining \eqref{Equ:SE2-a} and \eqref{Equ:SE2-b}, we have
\begin{subequations}
\begin{align}
&\frac{1}{{\left( {\frac{N}{{N - M}} \cdot v_{{\text{TSR-DFT}}}^{t + 1} } \right)^{ - 1}  + \left( {\frac{N}{M} \cdot v_{{\text{TSR-DFT}}}^t  + \frac{N}{M} \cdot \sigma ^2 } \right)^{ - 1} }}
\label{Equ:35a} \\
 &= \mr{mmse}\left( {\left( {\frac{N}{M} v_{{\text{TSR-DFT}}}^t  + \frac{N}{M}  \sigma ^2 } \right)^{ - 1} } \right).
\end{align}
\end{subequations}
From \eqref{Equ:monotone} and \eqref{Equ:norm} we have
\begin{equation}\label{Equ:36}
v_{{\text{TSR-DFT}}}^t  \ge v_{{\text{TSR-DFT}}}^{t + 1} .
\end{equation}
Replacing $v^{t}$ by $v^{t+1}$ in \eqref{Equ:35a}, and using \eqref{Equ:36}, we obtain the following inequality
\begin{subequations}
\begin{align}
&\frac{1}{{\left( {\frac{N}{{N - M}} \cdot v_{{\text{TSR-DFT}}}^{t + 1} } \right)^{ - 1}  + \left( {\frac{N}{M} \cdot v_{{\text{TSR-DFT}}}^{t + 1}  + \frac{N}{M} \cdot \sigma ^2 } \right)^{ - 1} }} \label{Equ:37a}
\\
 & \le \frac{1}{{\left( {\frac{N}{{N - M}} \cdot v_{{\text{TSR-DFT}}}^{t + 1} } \right)^{ - 1}  + \left( {\frac{N}{M} \cdot v_{{\text{TSR-DFT}}}^t  + \frac{N}{M} \cdot \sigma ^2 } \right)^{ - 1} }}
\\
& = \text{mmse}\left( {\left( {\frac{N}{M}  v_{{\text{TSR-DFT}}}^t  + \frac{N}{M} \sigma ^2 } \right)^{ - 1} } \right).
\end{align}
\end{subequations}
After some manipulations of \eqref{Equ:37a}, we get
\begin{subequations}\label{Equ:38}
\begin{align}
v_{{\text{TSR-DFT}}}^{t + 1}  + \frac{{\sigma ^2  \cdot v_{{\text{TSR-DFT}}}^{t + 1} }}{{\frac{N}{M} \cdot v_{{\text{TSR-DFT}}}^{t + 1}  + \frac{{N - M}}{M} \cdot \sigma ^2 }}\\
 \le \mr{mmse}\left( {\left( {\frac{N}{M} \cdot v_{{\text{TSR-DFT}}}^t  + \frac{N}{M} \cdot \sigma ^2 } \right)^{ - 1} } \right).
\end{align}
\end{subequations}
From \eqref{Equ:38} and noting the fact that $v^{t+1}_{\text{TSR-DFT}}\ge0$ (from \eqref{Equ:monotone} and \eqref{Equ:norm}), we have
\begin{equation}\label{Equ:39}
v_{{\text{TSR-DFT}}}^{t + 1}  \le \mr{mmse}\left( {\left( {\frac{N}{M} \cdot v_{{\text{TSR-DFT}}}^t  + \frac{N}{M} \cdot \sigma ^2 } \right)^{ - 1} } \right).
\end{equation}
Now consider AMP-IID. Combining \eqref{Equ:SE-AMP1} and \eqref{Equ:SE-AMP2}, we have
\begin{equation}\label{Equ:40}
v_{{\text{AMP-IID}}}^{t + 1}  = \mr{mmse}\left( {\left( {\frac{N}{M} \cdot v_{{\text{AMP-IID}}}^t  + \frac{N}{M} \cdot \sigma ^2 } \right)^{ - 1} } \right).
\end{equation}
Note that $\mr{mmse}(\cdot)$ is a monotonically decreasing function. Comparing \eqref{Equ:39} and \eqref{Equ:40} and based on the assumption that $v^t_{\text{TSR-DFT}}\le v^{t}_{\text{AMP-IID}}$, we readily obtain $v^{t+1}_{\text{TSR-DFT}}\le v^{t+1}_{\text{AMP-IID}}$, which proves \eqref{Equ:induct_proof}.
\end{IEEEproof}\par\vspace{3pt}
The MSE performances of TSR-DFT and AMP-IID at iteration $t$ are characterized by $\mr{mmse}\left( {\eta _{{\text{TSR-DFT}}}^{t + 1} } \right)$ and $\mr{mmse}\left( {\eta _{{\text{AMP-IID}}}^{t + 1} } \right)$, respectively. Corollary 1 below shows that TSR-DFT outperforms AMP-IID in terms of estimation MSE in each iteration.\vspace{3pt}
\begin{corollary}
$\mr{mmse}\left( {\eta _{{\text{TSR-DFT}}}^{t + 1} } \right) \le \mr{mmse}\left( {\eta _{{\text{AMP-IID}}}^{t + 1} } \right)$.
\end{corollary}\vspace{3pt}
\begin{IEEEproof}
By comparing \eqref{Equ:SE-AMP1} and \eqref{Equ:SE2-a}, together with Proposition 2, it is straightforward to see that $\eta _{{\text{TSR-DFT}}}^{t + 1}  \ge \eta _{{\text{AMP-IID}}}^{t + 1}$. Corollary 1 follows since $\mr{mmse}(\cdot)$ is a monotonically decreasing function.
\end{IEEEproof}
\section{Numerical Examples}
\begin{figure}[htbp]
\centering
\includegraphics[width=.45\textwidth]{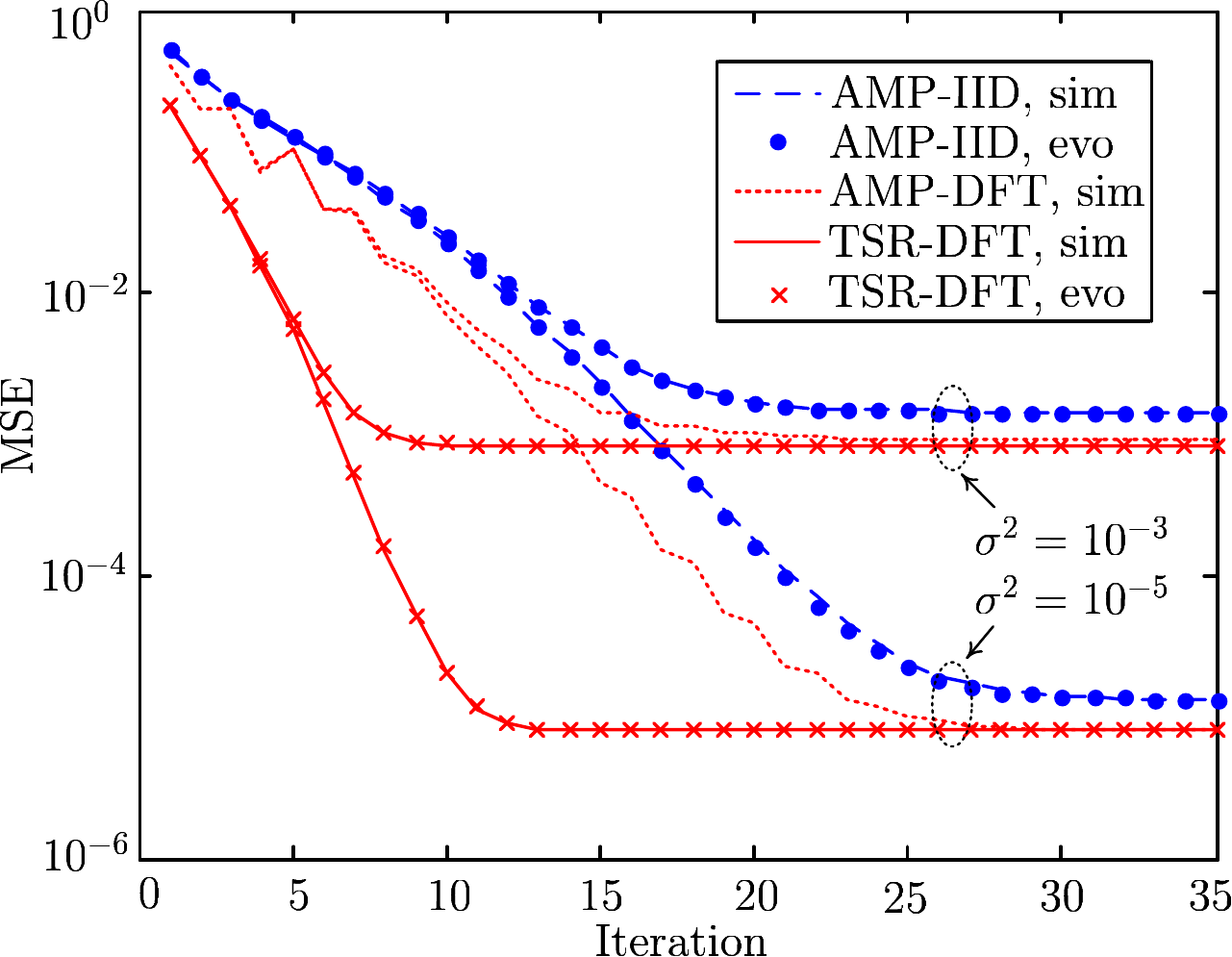}
\caption{MSE performances of TSR and AMP. $\lambda=0.4$. $N = 8192$. $M=5734\, (\approx 0.7N)$. The simulated MSEs are obtained by averaging over 500 realizations.}\label{Fig:numerical}
\end{figure}

Fig.~\ref{Fig:numerical} shows the numerical results for AMP-IID, AMP-DFT and TSR-DFT. First, we see that the simulation and evolution results for TSR-DFT and AMP-IID agree very well. Note that only simulation results are provided for AMP-DFT since no efficient analysis technique is available.\par
{{From Fig.~\ref{Fig:numerical}, we see that TSR-DFT outperforms AMP-IID in terms of both convergence speed and convergent MSE, which verifies Corollary 1. Also, the simulation results show that TSR-DFT converges faster than AMP-DFT. From Fig.~\ref{Fig:numerical}, it seems that the differences in the convergent MSEs are minor for TSR-DFT and AMP-DFT. However, if we decrease $M$, a more significant gain of TSR-DFT over AMP-DFT could be observed, see \cite[Fig.~3]{Ma2015}. \par
In simulations, we find that the performance advantage of TSR over AMP shrinks as $\lambda$ decreases. We will not show the results here due to space limitation.}}
\section{Conclusions}
{{In this letter, we proved based on state evolution that TSR-DFT outperformed AMP-IID. In addition, our simulation results suggest that TSR-DFT converges faster than AMP-DFT. Possible future work includes extending the TSR algorithm to the IID setting and compare it with AMP-IID.}}

\ifCLASSOPTIONcaptionsoff
  \newpage
\fi
\bibliographystyle{IEEEtran}	
\bibliography{IEEEabrv,CompressedSensing2}		

\end{document}